\documentclass{article}

\usepackage{ucs}
\usepackage[utf8x]{inputenc}
\usepackage[LGR,T2A,T1]{autofe}

\usepackage{graphicx}
\usepackage{url}

\begin{document}

\title{Java Code Analysis and Transformation into AWS Lambda Functions}
\author{Josef Spillner and Serhii Dorodko\\
Zurich University of Applied Sciences, School of Engineering\\
Service Prototyping Lab (blog.zhaw.ch/icclab/)\\
8401 Winterthur, Switzerland\\
\{josef.spillner,dord\}@zhaw.ch}

\maketitle

\begin{abstract}
Software developers are faced with the issue of either adapting their programming
model to the execution model (e.g. cloud platforms) or finding appropriate
tools to adapt the model and code automatically. A recent execution model which would benefit
from automated enablement is Function-as-a-Service.
Automating this process requires a pipeline which includes steps for code analysis,
transformation and deployment.
In this paper, we outline the design and runtime characteristics of Podilizer, a tool which
implements the pipeline specifically for Java source code
as input and AWS Lambda as output.
We contribute technical and economic metrics about this concrete 'FaaSification' process
by observing the behaviour of Podilizer with two representative Java software projects.
\end{abstract}


\section{Motivation}

Cloud computing brings along new programming models. Ranging from monolithic virtual machines
or containers over composite microservices to finer-grained functions and stream operators,
an application developer is faced with a complex choice of the right model and technology for cloud-enabled, cloud-aware
and cloud-native applications \cite{smemigration}. Once the model is chosen,
the switch to another model becomes non-trivial, in particular for the direction from coarse-
to fine-grained structures. As the Function-as-a-Service (FaaS) class is becoming more popular \cite{faastrends},
its exploitation would benefit from an automated transformation of legacy code and of
generic new code to the code conventions expected by this model. Program transformation and especially
source code transformation are established software engineering techniques for automated
profiling, security improvements, optimisation and refactoring \cite{spoon,jtransformer}.
In the quickly evolving cloud environments, it becomes essential to add exposure
to new target platforms to this list considering that many applications are merely programmed against specific interfaces
rather than generated through a top-down model-driven architecture \cite{mdacritique}.

While most FaaS providers support multiple programming languages, the scope of this study mandates
to pick one for a detailed analysis.
The Java programming language is widely taught and used to implement services. Typical programming models
supported by the default development kit encompass Servlets, EJBs and diverse component models, and JAX-WS/JAX-RS for web services.
Third-party Java frameworks offer even more choices, including Restlet and Apache CXF.
Therefore, despite being only supported by few FaaS providers, we pick Java as input format for our research.
The implications range beyond just programming. Most Java software projects are expected to be built
either manually by opening the project in a suitable development environment (e.g. Eclipse) or in
an automated way by executing a build script (e.g. Ant, Maven, Gradle). The build instructions
can be exploited to automate the software transformation.
On the provider side, we limit our study to Amazon Web Services (AWS) Lambda as it is one of the few services for
hosting Java functions as a service.

Hence, we aim at contributing novel insight into the automated transformation of legacy applications
into FaaS-hosted cloud applications which we name FaaSification. We claim that our tool,
Podilizer\footnote{Podilizer is publicly available at \url{https://github.com/serviceprototypinglab/podilizer}}, is the first one which
performs such a transformation from monolithic Java code to AWS Lambda units, and it does so with sufficient quality to be considered in cloud application
prototyping projects. Due to the restriction to Lambda, we refer to this process more specifically as Lambdafication.

Related work is available on the comparison of pricing effects for monolithic and microservice
applications using AWS Lambda \cite{lambdacost,microservicecosts}.
Compared to these general analyses and formalisations, this work motivates finding an automated transformation
to let developers make the policy choice at a late point in time while hiding the actual mechanism to enact the policy.
Further related work covers FaaS providers and implementations, for instance OpenLambda \cite{serverless}, but does not
contribute to their integration into the software and service engineering process.

The paper is structured as follows. First, we outline our research questions and approach and inform about how
to map object-oriented programming to functional services. The mapping description is complemented
by a an abstract pipeline architecture and a concrete architecture of our realisation thereof, the Podilizer tool.
Afterwards, we explain the experimental evaluation and discuss the extracted findings.
The paper concludes with an open discussion of how software should be written for the cloud.


\section{Approach}

Our approach consists of three parts. First, we identify two research questions. Then, we explain
general decisions which must be taken by any transformation tool related to the programming model,
the handling of stateful objects and the design of a transformation process.
The third part presents both the design and the implementation of our transformation tool Podilizer.

\subsection{Research Questions}

The planned code transformation process leads to two research questions ($RQ_1$ and $RQ_2$). Our approach is focused
on generating empirical results and deriving the answers accordingly.

$RQ_1$: Is it economically viable to run a Java application entirely over FaaS? The comparison baseline would be
conventional programmable platform (PaaS) models by deploying the application onto a suitable application server as well as
programmable infrastructure (IaaS) by wrapping the Java application into a container or a virtual machine.

$RQ_2$: Is it technically feasible to automate this process? And if so, which percentage of code coverage
can be expected, which performance can be achieved, and which code is easier, hard or impossible to convert?

\subsection{Programming $\mapsto$ Execution Model Mapping}

FaaS is inherently bound to the functional programming paradigm. Its characteristics under a pure interpretation
are determined by stateless computations with strict use of invocation parameters and return values without global variables.
In practice, just as functional programming languages have introduced techniques to cause side effects and manage state,
for instance through monads, so do most FaaS interfaces, for instance through access to storage services.

The function orientation is in contrast to Java's model which is predominantly an object-oriented language.
Even though few functional programming concepts are available
through the Functional Java library and starting with Java 8 with native Lambda expressions \cite{javalambdas}, the dominant
share of code is written in a pure object-oriented way. The same observation applies to similar programming language.
Therefore, the code translation needs to take the paradigm shift into account.
According to Plumbr, an application monitoring provider, the Java language versions in use in 2015
were Java 8 (20.84\%), Java 7 (59.37\%) and Java 6 (19.79\%) \cite{plumbr}. Following industry relevance,
our research focuses on Java 7/8.

The challenges are then related to the mapping of Java classes to appropriately packaged Java FaaS functions,
called FaaS units or functional units.
The mapping needs to account for empty methods, getters and setters, constructors and singletons.
Beyond the code, typical Java project conventions such as the presence of a \texttt{src} folder, but also
the absence thereof and exceptions from the conventions, need to be accounted for.
Finally, the mapping needs to consider the grouping of methods per functional unit to avoid
excessive network calls and ensure that all dependency methods referenced from each method can be resolved.

\subsection{State Handling}

Java methods are often stateful through instance attributes. The state handling of the resulting
decomposed functions can either extend the method signature to pass in and out all attributes
which are accessed and modified, or use server-side state.
AWS Lambda offers both an S3 blob storage interface and local environment variables. However,
the variables are restricted to read-only access from the runtime \cite{cloudcontrolplane}. Weighting
the advantages of S3 (performance) against extended method signatures (price, functional purity), our
approach uses the latter technique.

In Java, methods with parameters are integral parts of classes and are used to change the state of the corresponding
instances: $Class.method(params)$. The instance is self-referenced implicitly with the keyword $this$. According to the Lambda programming model, every function unit assumes a specific stateless
class with a method handler which is triggered when the function is invoked. The statelessness is due to not guaranteeing
the same object of this class to be used for subsequent requests. Early works to compile Java code into
a typed Lambda calculus have suggested making the self-references explicit by enhancing the method
signatures with it \cite{typedlambda} which is the approach chosen by us as well.

The translation process thus rewrites the method header with the Lambda-required signature and the method body with generated code.
This code first initialises the invocation credentials, creates an input object to save the instance state as well
as any method parameters,
initialises the Lambda invoker with the created input object serialised to JSON, calls
the method ($Class.ha\-ndleRequest(input,output,context)$), fetches the result from the deserialised output object,
and renews the instance
state using the result object. The credentials are read from the environment and upon failure
from a configuration file which permits the generated code to still run outside of the Lambda environment.

\subsection{FaaSification Pipeline}

\begin{figure}[h]
\center
\includegraphics[width=0.8\columnwidth]{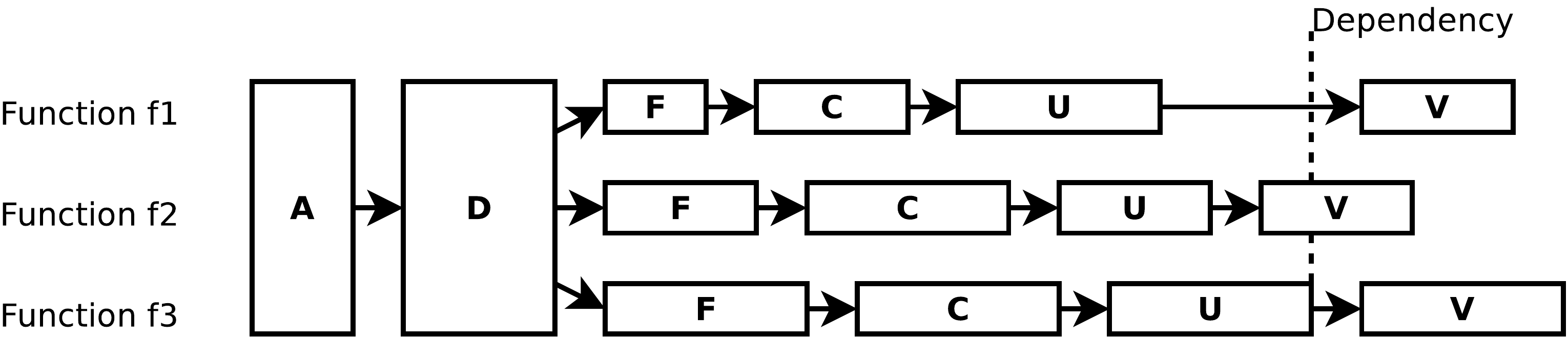}
\caption{FaaSification pipeline}
\label{fig:pipeline}
\end{figure}

FaaSification is the process of converting a code structure into a format which is executable on FaaS.
In our approach, this process is represented by a superscalar pipeline which allows for parallel
processing of each of its steps. The first step is the static code parsing and analysis ($A$). The second step
is the decomposition into functional units and a remainder which includes the code identified as
incompatible to the target FaaS platform ($D$). The third step is the source-to-source translation of the functional units into
FaaS units, adhering to the calling conventions of the target platform ($F$). The fourth step is the
compilation and dependency assembling of these units ($C$), and the fifth step is their upload, deployment
and configuration to turn them into ready to use microservices ($U$). An optional sixth step is the systematic
test of all deployed functions and the verification of the successful transformation ($U$).
Fig. \ref{fig:pipeline} gives an example of a FaaSification pipeline whose parallel execution depends on
resource consumption superpositioning and on dependencies between methods before the ability to run
unit tests.

\subsection{Podilizer Design}

The pipeline thinking is reflected in the design for a local tool called Podilizer run by software engineers as part of their
development environment. Podilizer implements the FaaSification pipeline in its Lambdafication flavour by recursively scanning directories
for Java projects and processing each project and each Java source file until the code is available to be invoked
as Lambda function. The pipeline steps are incremental
and the tool allows for continuations starting from each preceding step. This design choice makes it fault-tolerant
and debug-friendly in addition to an easy parallelisation.
Table \ref{tab:continuations} informs about the steps ($A→D→F→C→U→V$) and the associated continuation points
and requirements. The source build files are not used because the project is immediately dissected into functions.
The Maven build file per target function is generated by Podilizer but may need project-specific customisation
to incorporate dependency libraries into the build process. The AWS credentials are assumed to be readily
available on the local system, for instance by a prior installation of the AWS CLI. The remaining requirements
are all extracted from the analysed projects.

\begin{table}[htb]
\centering
\caption{FaaSification pipeline steps and continuations.\label{tab:continuations}}
\begin{tabular}{|l|l|l|} \hline
\textbf{Step}		& \textbf{Continuation Points}		& \textbf{Requirements}	\\ \hline

A: code analysis	& -- (internal AST)			& source code directory	\\ \hline
D: code decomposition	& -- (internal list of ASTs)		& --			\\ \hline
F: function translation	& target source directory		& --			\\ \hline
C: compilation		& target binary directory		& Maven buildfile (cust.)\\ \hline
U: upload		& deployed functions list		& AWS credentials	\\ \hline
V: verification		& --					& unit test definitions	\\ \hline
\end{tabular}
\end{table}

\subsection{Podilizer Implementation}

Podilizer is itself implemented in Java, using the JavaParser framework for static code analysis
and the Maven library to inject build instructions into the target software projects.
It executes the AWS CLI tool as external process for all interaction with AWS Lambda and any supported
unit test framework it finds for the verification step. Currently, JUnit is supported.

\begin{figure}[h]
\center
\includegraphics[width=0.8\columnwidth]{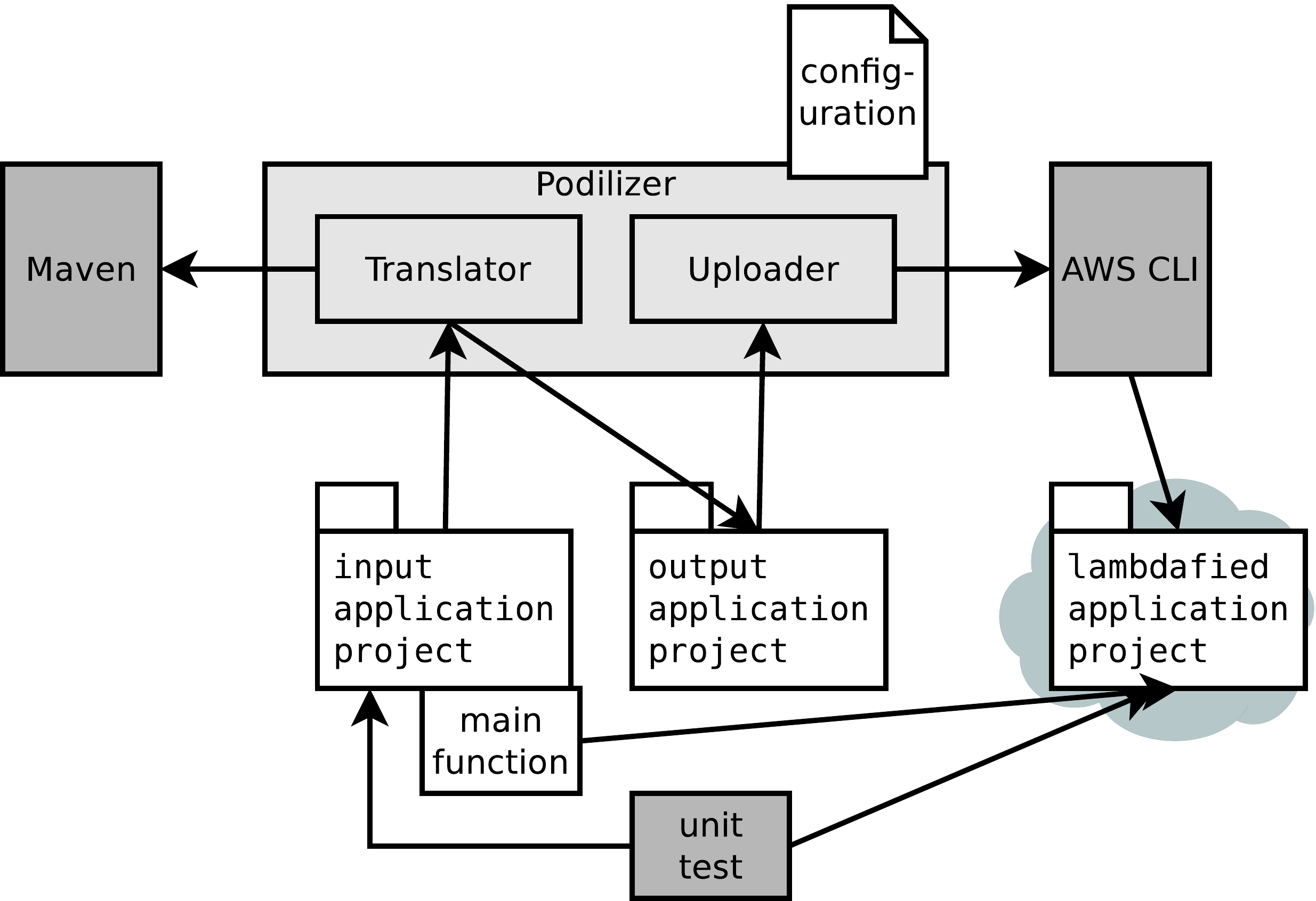}
\caption{Implementation architecture of Podilizer}
\label{fig:architecture}
\end{figure}

Fig. \ref{fig:architecture} gives an overview about the Podilizer implementation. Its main components
are the translator which covers the first half of the pipeline ($A→D→F[→C]$) in which the use
of Maven on all generated build files ($C$) is optional, and the triggering of the upload ($U$)
as well as the unit test execution ($V$).

The implementation complexity of Podilizer is moderate with about 2100 lines of Python code.
The invocation takes the most important parameters such as pipeline steps, source directory and target
directory as command-line parameters. Further configuration details can be specified in a
YAML file which is parsed on startup.


\section{Trials and Findings}

\begin{figure}[h]
\center
\includegraphics[width=0.8\columnwidth]{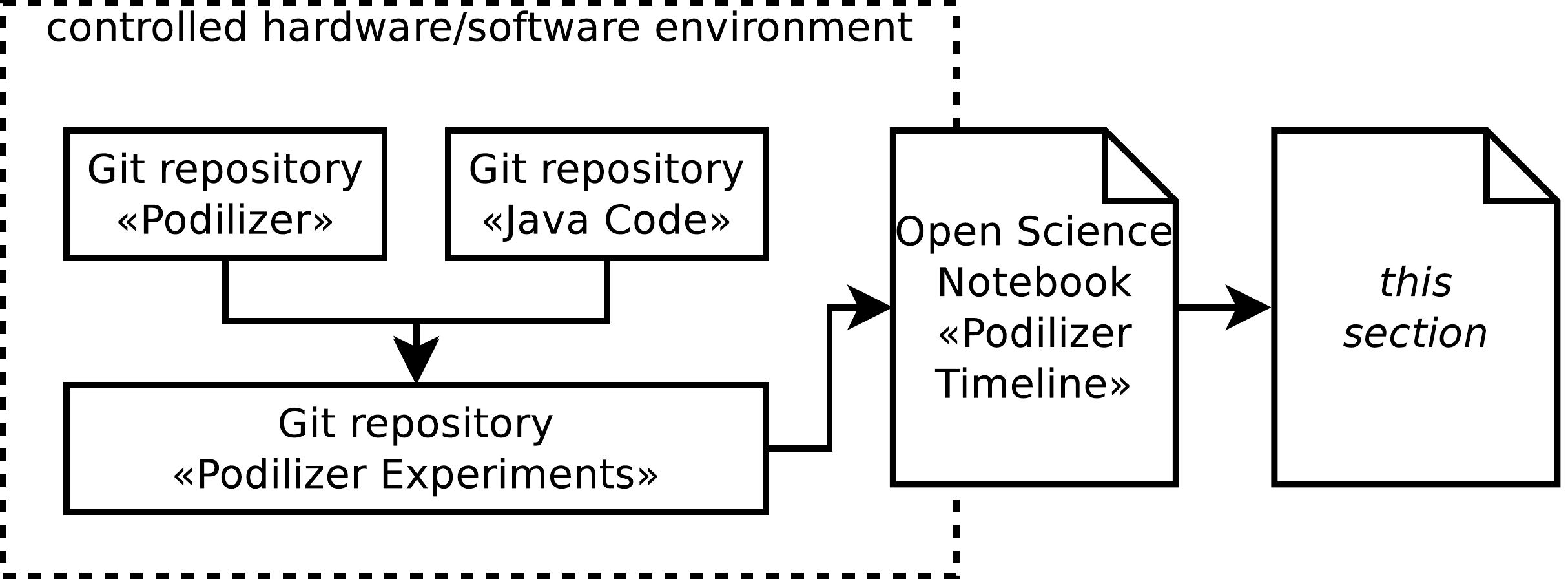}
\caption{Testbed for performing experiments on Podilizer}
\label{fig:testbed}
\end{figure}

We evaluate Podilizer experimentally with a standard research testbed approach shown in Fig. \ref{fig:testbed}.
All input to the experiments is recorded in public versioned repositories, and all raw output is captured
in another dedicated repository. The versioning allows for finding improvements and regressions over time
as the software evolves.
All experiments are tracked in a public Open Science Notebook and tools for reproducibility, repeatability and recomputation
are made available as well in a Podilizer Repeatability project\footnote{Podilizer Repeatability in the Open Science Framework: \url{https://osf.io/c886p/}}.
The main results of the current implementation are reflected in this section.

\subsection{Experiment Setup}

Each step of the pipeline is associated to a unique check for success.
The first three steps are performed internally by Podilizer within one procedure, whereas the three remaining ones are
merely automated by running executables out of which one is provided by Podilizer, too.
The most crucial check is the final one which is successful if all deployed functions are remotely invocable.
According to DZone, about 30.7\% of all Java projects hosted on Github depend on JUnit which calls for
an integration to ensure systematic testing of the deployment \cite{dzone}.
Table \ref{tab:pipeline} summarises all steps and checks.

\begin{table}[htb]
\centering
\caption{FaaSification pipeline steps and checks.\label{tab:pipeline}}
\begin{tabular}{|l|l|} \hline
\textbf{Step}		& \textbf{Check}			\\ \hline

A: code analysis	& JavaParser internal return value	\\ \hline
D: code decomposition	& Podilizer internal			\\ \hline
F: function translation	& Podilizer internal			\\ \hline
C: compilation		& compiler/build tool exit status	\\ \hline
U: upload		& Podilizer deployer exit status	\\ \hline
V: verification		& call test, unit test exit status	\\ \hline
\end{tabular}
\end{table}

Podilizer is instrumented with millisecond-precision logging to reveal the duration of each pipeline
step. In addition to the performance, the quality of the transformation can be measured by the ratio
of successful checks against all which are performed in each step.

The economic aspect requires a comparison between the execution of the lambdafied application compared
to a monolithic execution in a configuration which matches the performance. In the absence of a general performance
estimation formula, a manual calibration specific to each software application under test is therefore needed.

The reference input project set consists of six software applications which represent the large variety of Java software engineering,
ranging from 28 to 771 significant lines of code (SLOC), diverse interaction forms (none, standard input and output, graphical, files, HTTP methods)
and build tools (javac, make, maven, ant) and artefact type (applications, libraries, plugins, tests).
The software projects are a graphical window with buttons (P1), mathematical functions (P2), calculation
of shipping containers and boxes (P3), public transport information (P4), image processing (P5) and domain-specific
language parsing and evaluation (P6).

\subsection{Results}

We have run the experiment on a Dell Latitude E7450 notebook with Intel Core i7-5600 quad-core processor clocked at 2.60 GHz.
The notebook was connected to SWITCHlan, the Swiss university network, via 1000baseT Ethernet, and installed with
Ubuntu Linux and OpenJDK 8.
The results differ depending on the chosen software project to translate.
The values are also
influenced by the hardware, the used software tools (Podilizer, Maven, JUnit), the provider (AWS Lambda) and the network connection
in between. A full specification and self-contained virtual machine is made available as part of the Podilizer Repeatability
project.

Table \ref{tab:performancequalityp1} informs about the performance and quality of the FaaSification pipeline for project P1.
The values for $C$ and $D$ represent the sum of individual measurements for five functional units with a relatively
low deviation for local compilation but a higher one for network-dependent upload: $C_{min} = 3832ms$ and $C_{max} = 4038ms$,
$D_{min} = 5743ms$ and $D_{max} = 10414ms$.
The verification step is omitted due to the lack of unit tests in P1.

\begin{table}[htb]
\centering
\caption{FaaSification pipeline characteristics for P1.\label{tab:performancequalityp1}}
\begin{tabular}{|l|r|r|} \hline
\textbf{Step}		& \textbf{Performance}	& \textbf{Quality}	\\ \hline

A: code analysis	&  0.055s		& 100\%			\\ \hline
D: code decomposition	&  0.002s		& 100\%			\\ \hline
F: function translation	&  0.122s		& 100\%			\\ \hline
C: compilation		& 10.173s		& 100\%			\\ \hline
U: upload		& 21.238s		& 100\%			\\ \hline
V: verification		& --			& --			\\ \hline
TOTAL			& 31.590s		& success		\\ \hline
\end{tabular}
\end{table}

For comparison, Tables \ref{tab:performancep2} and \ref{tab:qualityp2} contain the separate performance and quality values for P2--P6.
Concerning the performance, the first two steps ($A$, $D$) almost always execute in less than a single Lambda billing period
($100ms$) whereas the functional decomposition takes up to one second depending on the code complexity.
The compilation and upload take consistently much longer in comparison. A just-in-time transformation
is precluded and optimisation techniques are needed to overcome this limitation. Nevertheless, the entire
transformation process including systematic unit testing performs in an acceptable timeframe and can be further
optimised by stronger parallelisation depending on the build system and unit test framework.

\begin{table}[htb]
\centering
\caption{FaaSification pipeline characteristics (performance) for P2--P6.\label{tab:performancep2}}
\begin{tabular}{|l|r|r|r|r|r|} \hline
\textbf{Step}	& \textbf{P2:P}	& \textbf{P3:P}	& \textbf{P4:P}	& \textbf{P5:P}	& \textbf{P6:P}	\\ \hline

A		& 0.054s	& 0.058s	& 0.074s	& 0.074s	& 0.105s	\\ \hline
D		& 0.002s	& 0.005s	& 0.010s	& 0.011s	& 0.028s	\\ \hline
F		& 0.096s	& 0.302s	& 0.867s	& 0.025s	& 0.701s	\\ \hline
C		& 10.530s	& 17.777s	& 37.707s	& --		& 22.901s	\\ \hline
U		& 21.349s	& 31.141s	& 65.075s	& --		& 44.858s	\\ \hline
V		& 11.942s	& --		& 13.927s	& --		& --		\\ \hline
TOTAL		& 43.973s	& 49.283s	& 117.657s	& --		& 68.593s	\\ \hline
\end{tabular}
\end{table}

The achieved quality is binary as the only failed transformation process (P5) is due to a crash of the
transformator itself. A more graceful partial transformation by tainting problematic methods would help
to raise the total percentage above 0\%.

\begin{table}[htb]
\centering
\caption{FaaSification pipeline characteristics (quality) for P2--P6.\label{tab:qualityp2}}
\begin{tabular}{|l|r|r|r|r|r|} \hline
\textbf{Step}	& \textbf{P2:Q}	& \textbf{P3:Q}	& \textbf{P4:Q}	& \textbf{P5:Q}	& \textbf{P6:Q}	\\ \hline

A		& 100\%		& 100\%		& 100\%		& 100\%		& 100\%		\\ \hline
D		& 100\%		& 100\%		& 100\%		& 100\%		& 100\%		\\ \hline
F		& 100\%		& 100\%		& 100\%		& 0\%		& 100\%		\\ \hline
C		& 100\%		& 100\%		& 100\%		& 0\%		& 100\%		\\ \hline
U		& 100\%		& 100\%		& 100\%		& 0\%		& 100\%		\\ \hline
V		& 100\%		& --		& 100\%		& --		& --		\\ \hline
TOTAL		& success	& success	& success	& fail		& success	\\ \hline
\end{tabular}
\end{table}

These results contain data points which lead to the answer of $RQ_2$. The automated translation is feasible,
with high code coverage for simple but heterogeneous code projects. The failures in the experiment are due
to dynamic classloading for plugins and the insufficient handling of such constructs by the transformer.

When the deployment process is finished, the interest shifts to the execution performance of
each software application.
Table \ref{tab:execperformance} compares the monolithic execution locally and on an equivalent IaaS and PaaS
setup (AWS EC2 and Elastic Beanstalk, respectively) against the one with decomposed functions for all six analysed
software projects. The EC2 execution occurrs on a single-core Ubuntu node with 3.5 GB main memory and an SSD.
It uses Xinetd to trigger the execution from an opened TCP connection to Xinetd's ports 10001 to 10006 for P1 to P6.
The local and EC2 invocations use the startup sequence of the build system, for instance $mvn:exec$, which
causes additional delays.

\begin{table}[htb]
\centering
\caption{Application execution performance comparison.\label{tab:execperformance}}
\begin{tabular}{|l|r|r|r|r|r|r|} \hline
\textbf{Flavour}	& \textbf{P1:X}	& \textbf{P2:X}	& \textbf{P3:X}	& \textbf{P4:X}	& \textbf{P5:X}	& \textbf{P6:X}	\\ \hline

Notebook local		& --		& 0.71s		& 1.87s		& 1.25s		& 0.08s		& 0.13s		\\ \hline
AWS EC2	local		& --		& 1.18s		& 2.99s		& 1.92s		& 0.09s		& 0.18s		\\ \hline
AWS EC2	Xinetd		& --		& 1.16s		& 2.86s		& 1.57s		& 0.12s		& 0.22s		\\ \hline
AWS Beanstalk		& --		& 0.36s		& 0.36s		& 1.79s		& --		& 0.36s		\\ \hline
AWS Lambda		& --		& 8.77s		& 9.74s		& 12.20s	& --		& --		\\ \hline
\end{tabular}
\end{table}

As expected, P1 fails on the server due to being a graphical application, and its execution time depends on the user
in the other cases. Therefore it is excluded from the comparison.
P5 requires an interactive command-line interface and cannot be instrumented in a web environment or through
function calls alone.
P6 fails unexpectedly due to missing symbol files for the parser. The solution necessitates a concept to deploy
dependency files in addition to dependency code which is not part of our design.

The first observation beyond the exclusions is that the EC2 instance is consistently slower than the notebook which can be attributed to the rather
coarse-grained configuration options offered by IaaS providers. A second observation is that the two layers of indirection
through Xinetd and a wrapper shell scripts do often not cause a significantly higher execution time.

These results give an answer to $RQ_1$. While the applications perform slower by about an order of magnitude compared to
typical IaaS or PaaS deployments, the economic feasibility is still in range for services which are not permanently
invoked. A concrete example would be P2 whose $main$ method invokes the $sum$ method once with a memory consumption
of 34 MB. While the total execution time from the client is 8.77s, the Lambda execution takes only 1.67ms which leads to an effective billing period of only 100ms. Beyond the free tier, the associated cost for a million calls per month in the region $us-west2$ would be 20.8 US\$. The same workload could be delivered by an on-demand virtual machine in EC2 at a minimum of 4.39 US\$ (t2.nano). The fewer calls are forecasted, the more the Lambda pricing model becomes effective, with the cutting point at around 210000 calls per month.

Table \ref{tab:codesize} concludes the observations with a comparison of the source code size
before and after the transformation. Due to generated boilerplate code and local method
code duplication, the overhead relative to the original size becomes sometimes significant.

\begin{table}[htb]
\centering
\caption{Application source code size comparison.\label{tab:codesize}}
\begin{tabular}{|l|r|r|r|r|r|r|} \hline
\textbf{Flavour}	& \textbf{P1:S}	& \textbf{P2:S}	& \textbf{P3:S}	& \textbf{P4:S}	& \textbf{P5:S}	& \textbf{P6:S}	\\ \hline

Original		& 20 kb		& 32 kb		& 44 kb		& 40 kb		& 40 kb		& 96 kb		\\ \hline
Lambdafied		& 548 kb	& 12436 kb	& 988 kb	& 2960 kb	& --		& 1796 kb	\\ \hline
Overhead		& 2640\%	& 38763\%	& 2145\%	& 7300\%	& --		& 1771\%	\\ \hline
\end{tabular}
\end{table}


\section{Discussion}

Our findings in automated Java code to Lambda units transformation look promising for future
cloud application engineering.
The results are also beneficial to programming education where rather simple object-oriented applications
are in wide use and educators regularly struggle to keep up with new application hosting
formats and platform services.

Difficulties originate from code which is not prepared for individual function access. According to a recent
study, at least 20\% of Java methods are too accessible ($public$ instead of $protected$ or $private$) \cite{methodaccess}, while
for our work, they are sometimes too inaccessible, although the solution to both is the same: powerful refactoring
tools for software engineers. Further difficulties originate from interfacing with the Java virtual machine,
for instance through the classloader, and with the command-line interface, as well as with data access
with differing file paths.

Future work identified by limitations of our approach encompasses the handling of dynamic classloading,
server-side state handling, FaaSification beyond Java as input language and AWS Lambda as target service,
as well as optimisations for attribute and method dependencies.


\section*{Acknowledgements}

This research has been supported by an AWS in Education Research Grant which helped us to run our experiments on AWS Lambda as representative public commercial FaaS.

\bibliographystyle{unsrt}
\bibliography{javafaas}

\end{document}